\documentclass[12pt]{article}

 \usepackage{graphicx}
 \usepackage[latin1]{inputenc}
 \usepackage{xspace}
 \usepackage{latexsym}
 \usepackage{amstext}
 \usepackage{amsxtra}
 \usepackage{makeidx}

 \newcommand{\bs}{\boldsymbol}

 \title{Bose-Einstein condensates in fast rotation}

\begin{document}
 \normalsize
\noindent

\title{Bose-Einstein condensates in fast rotation}

\author{S. Stock, B. Battelier, V. Bretin, Z. Hadzibabic, and J. Dalibard
\\ Laboratoire Kastler Brossel\footnote{Unit\'e de Recherche
de l'Ecole normale sup\'erieure et de
l'Universit\'e Pierre et Marie Curie, associ\'ee au CNRS.}, 24 rue
Lhomond, 75005 Paris, France}
\date{\today}

\maketitle

\begin{abstract}
In this short review we present our recent results concerning the
rotation of atomic Bose-Einstein condensates confined in quadratic
or quartic potentials, and give an overview of the field. We first
describe the procedure used to set an atomic gas in rotation and
briefly discuss the physics of condensates containing a single
vortex line. We then address the regime of fast rotation in
harmonic traps, where the rotation frequency is close to the
trapping frequency. In this limit the Landau Level formalism is
well suited to describe the system. The problem of the
condensation temperature of a fast rotating gas is discussed, as
well as the equilibrium shape of the cloud and the structure of
the vortex lattice. Finally we review results obtained with a
quadratic + quartic potential, which allows to study a regime
where the rotation frequency is equal to or larger than the
harmonic trapping frequency.
\end{abstract}


The possibility to obtain quantum degenerate gases by a
combination of laser and evaporative cooling has opened several
new lines of research, at the border of atomic, statistical and
condensed matter physics (for a review, see e.g.
\cite{Cornell02,Ketterle02,Pethickbook,Stringaribook}). Among
them, the rotation of a Bose-Einstein condensate raises many
interesting problems with respect to the case of a classical
fluid. Since the condensate is described by a macroscopic wave
function $\psi(\bs r)=\sqrt{\rho(\bs r)}e^{i\phi(\bs r)}$, where
$\rho$ and $\phi$ are the spatial density and phase of the fluid,
there exist strong constraints on the velocity field of the
rotating gas. In a place where the spatial density is not zero,
this velocity field is given by $\bs v=\hbar \bs \nabla \phi/M$
($M$ is the particle mass), hence $\bs \nabla\times \bs v=0$. The
circulation of the velocity field  is quantized along any close
contour on which $\rho\neq 0$, and it is a multiple of $h/M$. The
rotation of the fluid is thus only possible through the nucleation
of quantized vortices \cite{Lifshitz,Donnelly91}, which are
singular points (in 2 dimensions) or lines (in 3 dimensions) of
vanishing density, and around which the phase $\phi$ varies by
multiples of $2\pi$. Vortices are universal objects which appear
in many macroscopic quantum systems, such as superconductors and
superfluid liquid helium.

Since the achievement of Bose-Einstein condensation in atomic
gases, many experimental and theoretical studies have been devoted
to vortices in these systems (\cite{Fetter01} and refs. therein).
A typical experiment is the following: One starts with a
condensate initially at rest, confined in an axisymmetric trap
(symmetry axis $z$), and stirs it by applying an elliptic
potential rotating at frequency $\Omega$ around $z$. For very
small values of $\Omega$ no angular momentum is transferred to the
condensate. Just above a critical value $\Omega_c$, a single
vortex is nucleated \cite{Madison00}. For a system in thermal
equilibrium, the existence of this critical frequency can be
viewed as a manifestation of superfluidity: for a slow enough
rotation frequency, the stirrer cannot drag the condensate and set
it in motion. For stirring frequencies notably larger than
$\Omega_c$, the number of vortices in the condensate $N_v$
increases, and values up to $N_v= 200$ have been obtained
experimentally \cite{Aboshaeer01,Engels02}.

In the following we will be mostly interested in the large vortex
number case, which is achieved by choosing $\Omega$ close to the
trapping frequency $\omega_\perp$ in the $xy$ plane. Note that in
a purely harmonic trap, one must keep $\Omega$ below
$\omega_\perp$; the centrifugal force otherwise exceeds the
trapping force and the gas is destabilized \cite{Rosenbusch02}.
The main features of the vortex assembly in this large $N_v$
regime are well known. The vortices form a triangular Abrikosov
lattice with a surface density $n_v=M\Omega/(\pi \hbar)$
\cite{Feynman}. When $\Omega \to \omega_\perp$, the radius of the
gas tends to infinity since the confinement by the trapping
potential is nearly balanced by the centrifugal force. Since the
surface density of vortices is constant ($\sim M\omega_\perp/(\pi
\hbar)$), the number of vortices $N_v$ also increases to
arbitrarily large values.

In principle the number of vortices can reach and even go beyond
the atom number $N$. For such a fast rotation, the description of
the system by a single macroscopic wave function is expected to
fail, and the ground state of the system should be strongly
correlated. Up to now this regime has been investigated only
theoretically
\cite{Cooper01,Paredes01,Sinova02,Reijnders02,Regnault03} and will
not be addressed here.

The outline of this short review is as follows. In
section~\ref{setrot} we present the mechanism used to set a
condensate in rotation. Then in section~\ref{singlevx} we briefly
review some experiments performed with a single vortex condensate.
In section~\ref{manyvxharm}, we focus on the fast rotation regime
and discuss some important features of this system, such as its
condensation temperature and its equilibrium shape. We use the
Landau Level approach, which makes a nice connection between this
physical problem and that of charged particles in a uniform
magnetic field. In section~\ref{sec:quartic} we turn to a
configuration that we recently investigated in our laboratory,
which consists in superimposing a trapping quartic potential onto
the usual quadratic one. This allows us to explore the rotation
regime $\Omega>\omega_\perp$, and we review some results obtained
for the vortex patterns in this regime. We conclude in
section~\ref{conclusion} by giving some perspectives of this
rapidly evolving field of research.

Note that due to the lack of space this paper does not attempt to
be a full review of the work that has been performed on rotating
quantum gases but will be subject to the following restrictions:
First we will only discuss studies on single component condensates
and refer the reader interested in rotating spinor condensates to
the recent experiments of the Boulder group (\cite{Schweikhard04b}
and references therein). Second we will illustrate our discussion
using mainly experimental results from our laboratory and only
give references to the achievements of other groups. Finally our
reference list focuses on papers published after 2001. For a
discussion of earlier articles see the detailed review paper by
Fetter and Svidzinski \cite{Fetter01}.

\section{Setting a BEC in rotation}
\label{setrot}

In order to nucleate vortices in a condensate, two classes of
methods have been used. The first one consists in imprinting on
the condensate the phase pattern $e^{i\theta}$ of the desired wave
function. It was successfully implemented experimentally by the
Boulder group for a two-component condensate \cite{Matthews99}. A
related scheme, based on the adiabatic inversion of the magnetic
field at the center of the magnetic trap, has been used at MIT
\cite{Leanhardt02}. The second method, which is used in our group,
consists in using a mechanical stirring of the condensate. In
order to do this, one can use the potential created by a laser
\cite{Madison00,Aboshaeer01} or by a magnetic field
\cite{Haljan01,Hodby01}. While the phase-imprinting method is well
suited for nucleating a single vortex, the stirring approach seems
to be more flexible and allows nucleation of a large number of
vortices.

Once the gas is rotating, a third method can be used to increase
the angular momentum per particle. It consists in eliminating
atoms with an angular momentum smaller than the average, so that
the remaining particles rotate at a larger angular speed. This
``evaporative spinup" method has been implemented in Boulder
\cite{Engels03}.

We now present the system that we have been using for rotation
experiments in our group. We use rubidium ($^{87}$Rb)
Bose-Einstein condensates produced in a cylindrically symmetric
Ioffe-Pritchard trap, with a frequency $\omega_\perp$ in the $xy$
plane and $\omega_z$ along the $z$ axis. The magnetic trapping
potential thus reads:
 \begin{equation}
V_{\rm mag}=\frac{1}{2}M\omega_\perp^2 (x^2+y^2)+
\frac{1}{2}M\omega_z^2 z^2\ .
 \label{vmag}
 \end{equation}
Typically $\omega_\perp \sim 10\, \omega_z$ in our experiments so
that the equilibrium shape of the condensate is an elongated
cigar. For $\omega_z/2\pi \sim 10$~Hz and $N\sim 3\times 10^5$
rubidium atoms in the trap, the length of the cigar is $100\,\mu$m
and its diameter is $10\,\mu$m.

We stir the condensate with a laser beam propagating along the $z$
axis. The beam has an anisotropic cross section and its eigenaxes
rotate at a frequency $\Omega$. The time-dependent potential
created by the laser beam can be written as
 \begin{equation}
V_{\rm stir}(t) =\frac{\epsilon}{2}M\omega_\perp^2 \left[
(x^2-y^2)\cos(2\Omega t)+2xy\sin(2\Omega t) \right]\ .
 \end{equation}
The parameter $\epsilon$ is a dimensionless measure of the
relative strength of the stirring and the magnetic potentials. In
practice we choose $\epsilon \sim$ 2--10 \%.

\begin{figure}
\centerline{\includegraphics{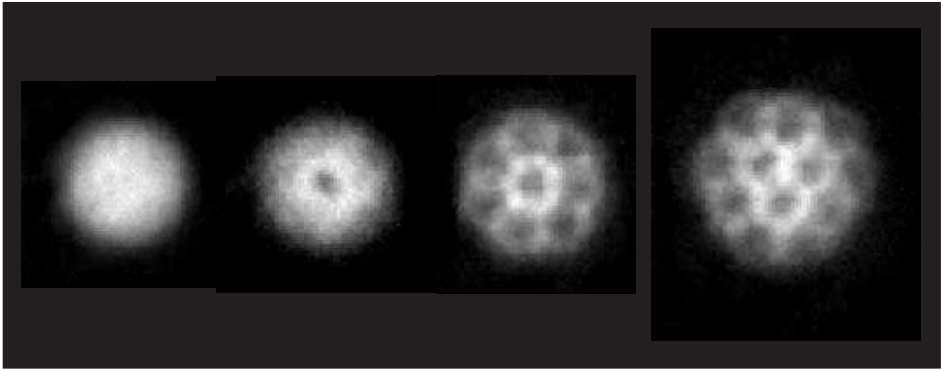}}
\caption{\emph{Quantized vortices.} Absorption images of a stirred
Rb Bose-Einstein condensate. The rotation frequency is increasing
from left to right (for details see \cite{Madison00}).}
\label{fig:images}
\end{figure}

We apply the stirring potential onto the condensate for a fraction
of a second, in order to transfer angular momentum to the gas. The
condensate then equilibrates in the cylindrically symmetric
potential (\ref{vmag}) for $\sim 1$ second. The trapping magnetic
field is then switched off and the gas undergoes a ballistic
expansion for a period of $\sim 20$~ms. Finally, we perform
absorption imaging along the rotation axis $z$. The vortices which
have been nucleated in this process appear as density dips in the
images, as seen in Fig.\ref{fig:images}.

When studying theoretically the problem of a rotating gas, one
usually assumes that an arbitrarily small anisotropic potential,
rotating at frequency $\Omega$, is added to the main isotropic
trapping potential. In presence of this stirring potential, the
frame rotating at $\Omega$ is the only one in which the state of
the system is stationary. The hamiltonian $H$ in this rotating
frame is deduced from the hamiltonian in the lab frame $H_{\rm
lab}$ by $H=H_{\rm lab}-\Omega L_z$. Experimentally, as we just
described, the rotating potential is often switched off for some
period before the measurement. $\Omega$ then plays the role of the
Lagrange multiplier associated with the deposited angular momentum
$L_z$, which is a constant of motion when the system evolves in
the axi-symmetric potential.

In the following we will mainly focus on the equilibrium
properties of the rotating system. We refer the reader interested
in the dynamics of vortex nucleation and decay to
\cite{Caradoc99,Sinha01,Recati01,Madison01,Dalfovo01,Linn01,Aboshaeer02,Williams02,Simula02,Tsubota02,Penckwitt02,Kasamatsu03,Lundh03,Lobo04,Duine04}.

\section{Single vortex physics}
\label{singlevx}

For a proper choice of $\Omega$ and the equilibration time after
stirring, it is possible to nucleate in a reliable way a single
vortex in the center of the condensate. Several experimental
studies have been performed on such a system. First, the average
angular momentum per particle $L_z$ has been measured and found to
be of the order of $\hbar$ \cite{Chevy00}. This measurement was
performed using the relation between $L_z$ and the frequencies of
the transverse quadrupole modes of the condensate
\cite{Zambelli98}. Complementary information has been obtained
using atom interferometry to measure the phase pattern of the wave
function \cite{Inouye01,Chevy01}.

\begin{figure}
\centerline{\includegraphics[width=10cm]{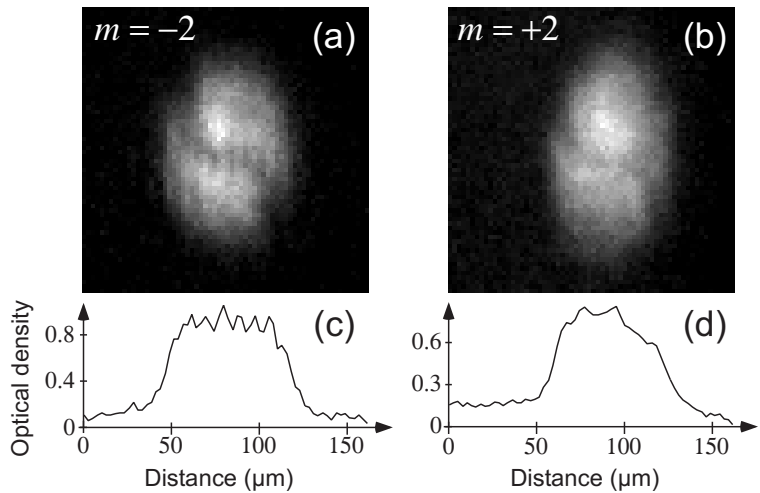}}
\caption{\emph{Kelvin mode of a vortex line}. Transverse images of
a Bose-Einstein condensate with a single, positively charged
vortex, obtained after excitation of the transverse quadrupole
mode $m=-2$ (a) and $m=+2$ (b). Fig.~\ref{fig:kelvons}a gives an
evidence for the Kelvin mode of the vortex line. This mode has an
angular momentum $m=-1$, and it is thus excited by the decay of
the transverse quadrupole mode $m=-2$ into two ``kelvons" (quanta
of the Kelvin mode). The two kelvons have the same energy and
propagate in opposite directions, which ensures the conservation
of linear momentum. By contrast, Fig.~\ref{fig:kelvons}b shows no
oscillation of the vortex line, as expected since the decay of the
quadrupole mode $m=+2$ into kelvons ($m=-1$) is forbidden by
angular momentum conservation. Figs. \ref{fig:kelvons}(c,d) are
the corresponding density profiles (for details see
\cite{Bretin03}).} \label{fig:kelvons}
\end{figure}

Concerning the vortex line itself, its equilibrium shape has been
determined: the line is often curved at the two ends of the cigar
\cite{Rosenbusch02b}. This bending is a symmetry breaking effect,
and it can be understood by noticing that a radially centered
vortex is favored at the axial center of the cigar, where the
density is large, whereas it costs less energy to radially
off-center the vortex line at the ends of the cigar, where the
density is low
\cite{Svidzinsky00,Feder01,Garcia01a,Garcia01b,Aftalion01,Modugno03}.
Some specific normal modes of the vortex line have also been
observed, such as the precession of a single vortex when it is not
aligned with the trap axes
\cite{Fedichev99,Anderson00,Stringari01,Hodby03}, and the Kelvin
mode of the vortex line (see Fig. \ref{fig:kelvons})
\cite{Bretin03,Mizushima03,Chevy03,Duine03,Fetter04}.

\section{Fast rotation in a harmonic potential}
\label{manyvxharm}

We now address the case of a fast rotating gas where the average
angular momentum per particle is large compared to $\hbar$, i.e.
in which many vortices have been nucleated. We consider that the
gas is confined in a purely harmonic potential (\ref{vmag}). We
first present the Landau level approach to this problem, which is
directly connected to the description of the motion of a charged
particle in a uniform magnetic field
\cite{Girvin84,Rokhsar99,Ho01}. We then address the determination
of the critical temperature of the rotating gas, and we compare
the result of the semi-classical approach given in
\cite{Stringari99} with the treatment using the Landau level
basis. We then turn to the discussion of the equilibrium shape of
the rotating condensate and the structure of the vortex pattern.

\subsection{The Landau level approach}

In the frame rotating at frequency $\Omega$, the non-interacting,
single particle hamiltonian is $H_0=H_\perp+H_z$ with
$H_z=P_z^2/(2M)+ M\omega_z^2 z^2/2$ and
 \begin{eqnarray}
H_\perp
&=&\frac{P_x^2+P_y^2}{2M}+\frac{1}{2}M\omega^2_\perp(x^2+y^2)-\Omega
L_z \label{singlepartH0}\\
&=&\frac{(\bs P_\perp-\bs
A)^2}{2M}+\frac{1}{2}M(\omega_\perp^2-\Omega^2) (x^2+y^2)\ ,
 \label{singlepartH}
 \end{eqnarray}
where the vector potential is $\bs A= M\bs \Omega\times \bs r$.
Eq. (\ref{singlepartH}) is formally identical to the hamiltonian
of a particle of charge 1 placed in a uniform magnetic field
$2M\Omega \bs {\hat z}$, and confined in a harmonic potential of
frequency $\sqrt{\omega_\perp^2-\Omega^2}$. The Coriolis force,
which has the same mathematical structure as the Lorentz force,
originates from the vector potential $\bs A$, whereas the term
$-M\Omega^2(x^2+y^2)/2$ corresponds to the centrifugal potential.

\begin{figure}
 \centerline{\includegraphics[width=7cm]{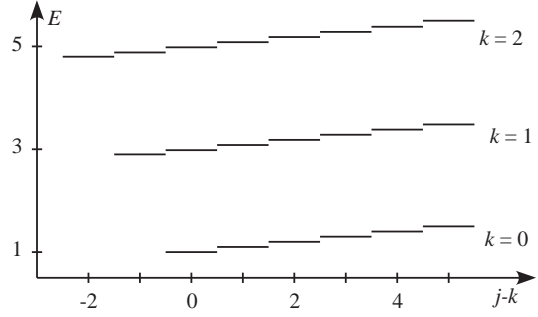}}
\caption{\emph{Landau level structure.} Single particle spectrum
for the transverse motion (in the $xy$ plane) for $\Omega=0.9$.
The index $k$ labels the Landau levels. The energy is expressed in
units of $\hbar \omega_\perp$. }
 \label{fig:LL}
\end{figure}

Common eigenstates of $H_0$ and $L_z$ have single particle
energies
 \begin{equation}
 E_{j,k,n}/\hbar= \omega_\perp+\frac{\omega_z}{2}
 +j(\omega_\perp-\Omega)+ k(\omega_\perp+\Omega)+ n\omega_z
\label{spectrum}
 \end{equation}
and angular momentum $\hbar(j-k)$, where $j,k,n$ are non-negative
integers. For $\Omega$ close to $\omega_\perp$, the contribution
of the transverse motion to these energy levels (terms in $j$ and
$k$) groups in series of states with a given $k$, corresponding to
the well known Landau levels (Fig. \ref{fig:LL}). The lowest
energy states of two adjacent Landau levels are separated by
$\hbar(\omega_\perp + \Omega)$, whereas the distance between two
adjacent states in a given Landau level is
$\hbar(\omega_\perp-\Omega)$. When $\Omega = \omega_\perp$, all
states in a given Landau level are degenerate. Physically, this
corresponds to the case where the centrifugal potential exactly
balances the trapping force in the $xy$ plane, and only the
Coriolis force remains. The system is thus invariant under
translation, hence the macroscopic degeneracy.

When interactions between particles are taken into account, the
Landau levels are no longer eigenstates of the $N-$body
hamiltonian. However they are still relevant in the regime of fast
rotation. Indeed as $\Omega\to \omega_\perp$, the restoring force
in the $xy$ plane becomes very small and the density of the gas
drops. The interaction energy per particle is then small compared
to the distance $2\hbar \omega_\perp$ between two Landau levels,
and the states of interest are essentially those associated with
$k=0$, i.e. the lowest Landau level (LLL)
\cite{Girvin84,Rokhsar99,Ho01}. Any function $\psi(x,y)$ of the
LLL can be cast in the form:
 \begin{equation}
\psi(x,y)=e^{-(x^2+y^2)/2a^2_\perp}\; P(x+iy)\ ,
 \label{formLLL1}
 \end{equation}
where $a_\perp^2=\hbar/(m\omega_\perp)$ and $P(u)$ is a polynomial
(or other analytic function) of $u$. When $P(u)$ is a polynomial
of degree $n$, it has $n$ complex zeroes. Each zero is the
position of a singly-charged, positive vortex, since the phase of
$\psi(\bs r)$ changes by $2\pi$ along a closed contour encircling
the zero.

\subsection{Critical temperature for a rotating gas}

The critical temperature for an ideal rotating gas in a harmonic
potential has been derived by S.~Stringari using a semi-classical
approach \cite{Stringari99}. Let us briefly outline the reasoning.
One starts from the semi-classical relation between the atom
number $N$ and the temperature $T$ at the BEC transition:
 \begin{equation}
N=\frac{1}{h^3}\int d^3r\;d^3p\; \frac{1}{\exp(H_0(\bs r,\bs
p)/k_BT)-1}\ ,
 \label{critical0}
 \end{equation}
where we assume that the minimum of the trapping potential is at
zero energy, so that we set the chemical potential also equal to
zero at the transition point. Using the form (\ref{singlepartH})
of $H_\perp$ and making the change of variables $p'_x=p_x+M\Omega
y$, $p'_y=p_y-M\Omega x$, we obtain a new integral. This integral
is identical to the one giving the condensation criterion for a
gas at rest, confined in a cylindrically symmetric, harmonic
potential. The transverse and longitudinal frequencies for this
model system are $\sqrt{\omega_\perp^2-\Omega^2}$ and $\omega_z$,
respectively, and we thus get:
 \begin{equation}
N=\zeta(3) \left(\frac{k_B T}{\hbar \bar \omega}\right)^3 \
,\qquad \bar \omega^3=(\omega_\perp^2-\Omega^2)\omega_z\ ,
 \label{critical}
 \end{equation}
where $\zeta(x)=\sum_n n^{-x}$ is the Riemann zeta function
($\zeta(3)\simeq 1.202$). This entails that, within the
semi-classical approximation, the Coriolis force associated with
the vector potential $\bs A(\bs r)$ has no effect on the critical
temperature. This is formally identical to the Bohr -- van Leeuven
theorem, stating that there is no magnetism at thermal equilibrium
in a system of charges described by classical mechanics. The only
effect of rotation in (\ref{critical}) is the change of the
transverse frequency due to the centrifugal force.

The result (\ref{critical}) can be recovered from the exact
one-body spectrum in terms of Landau levels given in
(\ref{spectrum}). Indeed this spectrum is the same as that of a 3D
harmonic oscillator with frequencies $\omega_\perp-\Omega$,
$\omega_\perp+\Omega$, $\omega_z$, whose geometrical mean is the
frequency $\bar \omega$, hence the result (\ref{critical}).

We now discuss briefly the validity condition of
(\ref{critical}).~From the reasoning based on the Landau level
structure, we see that $k_BT$ must be large compared to each of
the three energies $\hbar(\omega_\perp\pm \Omega)$ and $\hbar
\omega_z$. In practice, we have $\omega_\perp\gg\omega_z$ for a
cigar-shape trap, so that when $\Omega\sim \omega_\perp$, the most
stringent validity condition for the use of (\ref{critical}) is
$k_B T \gg 2\hbar \omega_\perp$. It is not easy to recover this
validity condition directly from the use of (\ref{critical}).
Indeed one could have naively expected that $k_B T\gg \hbar
\sqrt{\omega_\perp^2-\Omega^2}, \hbar \omega_z$ would be a
sufficient condition, which is clearly not the case.

\subsection{The equilibrium shape of the rotating condensate}

We now suppose that the gas is at zero temperature and review
possible approaches for determining its ground state when
repulsive interactions are taken into account. As usual in the
physics of cold gases, these interactions are assumed to be
point-like, and they are characterized by the s-wave scattering
length $a_s$. The ground state $\phi(\bs r)$ of the gas is
obtained in the mean-field approximation by minimizing the energy
per particle \cite{Pethickbook,Stringaribook}
 \begin{equation}
E[\phi]=\int \left( \phi^* \left[H_0\phi\right] + \frac{Ng}{2}
|\phi|^4  \right) \; d^3r
 \label{energy1}
 \end{equation}
where $g=4\pi \hbar^2 a_s/M$ and $\phi$ is normalized to unity.

\subsubsection{Rotational hydrodynamics approach}

This approach is based on the approximation of diffused vorticity,
where the singularities of the velocity field $\bs v(\bs r)$ and
of the atom density $n(\bs r)$ at each vortex core are averaged
out. This method is adequate for describing the system at
macroscopic distances, larger than the intervortex spacing. From
the equations of motion of rotational hydrodynamics
\cite{Chevy03,Cozzini03} (see also \cite{Baym04a}), one derives
the steady state velocity field $\bs v(\bs r)=\bs \Omega \times
\bs r$ and the spatial density $n(\bs r)=N|\phi(\bs r)|^2$:
 \begin{equation}
n(\bs r)=\frac{1}{g}\;\mbox{max} \left(\;0\;,\; \mu-V_{\rm
mag}(\bs r)+\frac{M\Omega^2}{2}(x^2+y^2)\;\right)
 \label{TF1}
 \end{equation}
where $\mu$ is the chemical potential. This density profile is the
usual inverted parabola corresponding to the Thomas-Fermi result,
for an axisymmetric potential with frequencies
$\sqrt{\omega_\perp^2-\Omega^2}$ in the $xy-$plane and $\omega_z$
along the $z$ axis.

The result (\ref{TF1}) is valid only if $\mu \gg \hbar \omega_z$.
In the opposite regime $\mu \ll \hbar \omega_z$, the $z$ motion is
``frozen" to its ground state (a gaussian of extension
$a_z=\sqrt{\hbar/(M\omega_z)}$~). The relevant wavefunctions can
be written as $\phi(\bs r)=\psi(x,y)\;e^{-z^2/2a_z^2}$ and one has
to minimize
 \begin{equation}
E_\perp[\psi]=\int \left( \psi^* \left[H_\perp\psi\right] +
\frac{NG}{2} |\psi|^4  \right) \; d^2r
 \label{energy2}
 \end{equation}
where $G=g/( \sqrt{2\pi}\,a_z)$. The spatial density in the
transverse plane $n(x,y)=N|\psi(x,y)|^2$ is then:
 \begin{equation}
n(x,y)= \frac{1}{G}\;\mbox{max} \left(\;0\;,\;\mu
-\frac{1}{2}M(\omega_\perp^2-\Omega^2)(x^2+y^2)\;\right)\ ,
 \label{TFRHA}
 \end{equation}
As for the determination of the critical temperature, only the
centrifugal potential is important in this approximation. The
Coriolis force plays no role in the global equilibrium shape of
the condensate.

\subsubsection{Equilibrium shape in the LLL}

We now suppose that the interaction strength is small enough so
that the ground state of the system is essentially a LLL wave
function, corresponding to the quantum number $k=0$ in
Eq.~\ref{spectrum} ($\mu \ll 2\hbar\omega_\perp$). We also assume
that $\mu \ll \hbar \omega_z$ so that the $z$ motion is ``frozen"
to its ground state ($n=0$ in Eq.~(\ref{spectrum})). The use of
LLL wave functions allows to notably simplify the energy
functional in Eq.~(\ref{energy2}). One can indeed prove after some
algebra the two equalities:
 \begin{equation}
\langle E_{\rm kin} \rangle=\langle E_{\rm ho} \rangle =
\frac{\hbar \omega_\perp}{2}+\frac{\omega_\perp}{2}\int \psi^*
\left[L_z\psi\right] \; d^2r
 \label{equalities}
 \end{equation}
where the kinetic and harmonic oscillator energies are:
\begin{equation}
\langle E_{\rm kin} \rangle = \frac{\hbar^2}{2M}\int  |\bs \nabla
\psi|^2 \; d^2r \qquad  \langle E_{\rm ho}
\rangle=\frac{M\omega_\perp^2}{2}\int r^2\, |\psi|^2 \; d^2r\ .
\end{equation}
The energy is then given by
 \begin{equation} E[\psi]=\hbar\Omega +
 \int \left( M\omega_\perp(\omega_\perp-\Omega) r^2 |\psi|^2 +
\frac{NG}{2} |\psi|^4  \right) \; d^2r\ .
 \label{energy3}
 \end{equation}
We can express the distances and the energies in units of
$a_\perp=\sqrt{\hbar/M\omega_\perp}$ and $\hbar \omega_\perp$,
respectively. We then find that the minimization depends only on
the dimensionless parameter \cite{Aftalion04}
 \[
 \Lambda=N \;
 \frac{MG}{\hbar^2}\;
 \frac{\omega_\perp}{\omega_\perp-\Omega}
 =\sqrt{8\pi}\;N\;\frac{a_s}{a_z}\;
 \frac{\omega_\perp}{\omega_\perp-\Omega}
 \]
When $\Lambda <1$, the interaction term $NG|\psi|^4$ plays a
negligible role and the minimizing function is essentially the
ground state of the one-body hamiltonian $j=k=n=0$. For $\Lambda
\gg 1$ the minimum energy state is a linear combination of several
states corresponding to different quantum numbers $j$'s, and it
involves several vortices in the region where the atomic density
is significant.

The minimization of (\ref{energy3}) within the LLL has recently
been discussed in \cite{Watanabe04,Cooper04,Aftalion04}. Let us
briefly sketch the main results. One first defines the
coarse-grain average $\bar n(x,y)$ of the spatial density
$n(x,y)=N|\psi(x,y)|^2$, in order to smooth the rapid variations
at the vortex cores. The energy functional (\ref{energy3}) can be
written in terms of $\bar n$ instead of $n$, provided the
interaction parameter $G$ is renormalized to $bG$, where $b\simeq
1.16$ is the so-called Abrikosov parameter \cite{Kleiner64}. This
parameter arises from the discreteness of the vortex distribution:
since the wave function $\psi(x,y)$ must vanish at the vortex
location, the average value of $|\psi|^4$ over the unit cell,
hence the interaction energy, is larger than the result obtained
if $|\psi|$ was quasi-uniform over the cell. Once this
renormalization of $G$ has been performed, the minimization can be
performed by letting $\bar n$ vary over the whole space of
normalisable functions, the only constraint being that $\bar n$
varies smoothly over $a_\perp$, which is the characteristic length
scale for vortex spacing. One finds that the coarse-grain average
of the spatial distribution is the inverted parabola~:
 \begin{equation}
\bar n(x,y)=\frac{1}{bG}\;\mbox{max}
\left(\;0\;,\;\mu-M\omega_\perp(\omega_\perp-\Omega)(x^2+y^2)\;\right)
 \label{ThomasFermiLLL2}
 \end{equation}
This result is valid when the chemical potential $\mu$ is much
smaller than the distance $2\hbar\omega_\perp$ between the LLL and
the first excited LL, which amounts to:
 \[
N \;\frac{a_s}{a_z} \ll \frac{\omega_\perp}{\omega_\perp-\Omega}
 \]
Except for the Abrikosov coefficient $b$, the two results
(\ref{TFRHA}) and (\ref{ThomasFermiLLL2}) nearly coincide in the
fast rotation limit, since $\omega_\perp^2-\Omega^2 \simeq
2\omega_\perp(\omega_\perp-\Omega)$ when $\Omega\sim\omega_\perp$.

The fact that the equilibrium shape of the condensate remains an
inverted parabola even when the dynamics is restricted to the LLL
has been checked experimentally by the Boulder group
\cite{Schweikhard04,Coddington04}.

\subsubsection{Structure of the vortex pattern}

In first approximation the surface density $n_v$ of vortices in a
fast rotating condensate is uniform. One can show in this case
that the coarse-grain average of the velocity field is equal to
the rigid-body rotation result $\bs v(\bs r)=\Omega \hat {\bs
z}\times \bs r$ \cite{Feynman}, with $\Omega=\pi \hbar n_v/M$. For
rotating BECs, this relation has been checked experimentally at
MIT \cite{Raman01}. The vortices form a triangular lattice which
is known to minimize the interaction energy $g\int |\psi|^4$
\cite{Castin99}.

\begin{figure}
\centerline{\includegraphics{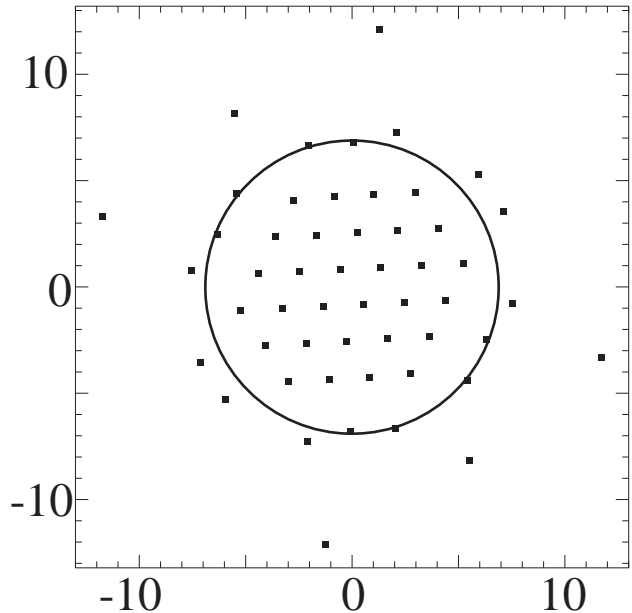}} \caption{\emph{Vortices
in the LLL.} Equilibrium vortex pattern obtained by minimizing the
energy within the LLL, for $\Lambda=3000$ (figure extracted from
\cite{Aftalion04}). The distances are expressed in units of
$a_\perp$. The LLL trial wave functions have 52 vortices and the
circle represents the border of the Thomas-Fermi distribution
(\ref{ThomasFermiLLL2}).} \label{vortexpattern}
\end{figure}

A closer analysis of the vortex distribution shows that the vortex
distribution is distorted on the edges of the condensate
\cite{Sheehy04,Watanabe04,Cooper04,Aftalion04}. The distortion is
particularly clear in the LLL, as it can be seen in
Fig.~\ref{vortexpattern} obtained by Aftalion et al.
\cite{Aftalion04}, where an example of vortex distribution is
given for the particular case $\Lambda=3000$. This distortion of
the vortex lattice is essential to ensure the proper decay of the
atomic density given in (\ref{ThomasFermiLLL2}). Indeed an LLL
wave function with a uniform vortex lattice always leads to a
Gaussian average distribution $\bar n (x,y)$ \cite{Ho01}, instead
of the predicted and observed inverted parabola
(\ref{ThomasFermiLLL2}).

Another interesting characteristic of the LLL is that the vortex
core is of the same size as the distance between adjacent vortices
($\sim a_{\bot}$). In this respect the entrance in the LLL for a
rotating condensate in a magnetic trap is the equivalent of the
field $H_{c2}$ in a type II superconductor
\cite{Fischer03,Baym04a}. Finally we note that the dynamics of the
vortex lattice itself raises many interesting problems. In
particular the so-called Tkachenko modes of the lattice have been
analyzed theoretically
\cite{Anglin02,Baym03,Choi03,Simula04,Mizushima04,Gifford04,Cozzini04}
and observed experimentally \cite{Coddington03}.

\section{Fast rotation in a quadratic+quartic potential}

\label{sec:quartic}

In this section we discuss some results obtained with a quartic
potential $\gamma r_\perp^4$ added to the usual harmonic
confinement (we set $r_\perp^2=x^2+y^2$). This quartic confinement
allows to study the regime of rotation where
$\Omega>\omega_\perp$. This regime is unreachable otherwise, since
the expelling centrifugal potential $-M\Omega_\perp^2 r_\perp^2$
would exceed the trapping potential. The properties of the
rotating gas in a quartic potential have recently attracted a lot
of theoretical attention
\cite{Fetter01b,Kasamatsu02,Lundh02,Kavoulakis03,Fetter03,Aftalion03,Jackson04,Fetter04b}.

\subsection{Implementation of a quartic potential}

Experimentally, we have created a quartic potential using a far
detuned laser beam (wavelength 532~nm), propagating along the axis
of the trap \cite{Bretin04}. The waist $w$ of the beam is larger
that the condensate radius, so that the potential created by the
laser $U_0\;\exp(-2r_\perp^2/w^2)$  can be written as:
 \begin{equation}
U(\bs r)\simeq U_0 - \frac{2U_0}{w^2}r_\perp^2
+\frac{2U_0}{w^4}r_\perp^4\ .
 \label{potquart}
 \end{equation}
The laser frequency is larger than the atom resonance frequency,
so $U_0>0$. The second term in (\ref{potquart}) leads to a
reduction of the transverse trapping frequency $\omega_\perp$ and
the third term provides the desired quartic confinement, with
$\gamma=2U_0/w^4$. In our experiments $\gamma=6.5\times
10^{-12}$~Jm$^{-4}$.

The total trapping potential in the $xy$ plane can be written as:
 \begin{equation}
V(\bs r_\perp)=\frac{1}{2}M\omega_\perp^2 r_\perp^2 \left(1+\tilde
\gamma\;\frac{r_\perp^2}{a_\perp^2}\right)\ ,
 \end{equation}
where the dimensionless number $\tilde \gamma=2\hbar \gamma
/(M^2\omega_\perp^3)$ characterizes the relative strength of the
quartic and the quadratic potentials. For our setup we have
$\omega_\perp/2\pi=65$~Hz and $\tilde \gamma\simeq 10^{-3}$. Hence
the quartic term is only a small perturbation of the ground state
of the one-body hamiltonian. Of course its importance grows when
one considers large $L_z$ states, in which the particle is
localized further away than $a_\perp$ from the center of the trap.

\subsection{Critical temperature}

We now determine the critical temperature $T_c$ for a gas rotating
in a quadratic+quartic potential. For simplicity we consider the
case $\Omega=\omega_\perp$ and we use the semiclassical
approximation, which is valid if $k_B T_c \gg \hbar \omega_\perp$.
Inserting
 \begin{equation}
H_0(\bs r, \bs p)=\frac{p^2}{2M} +\frac{1}{2}M\omega_z^2 z^2
+\gamma r_\perp^4
 \end{equation}
in (\ref{critical0}) we obtain
 \begin{equation}
N=\zeta(5/2)\,\frac{\sqrt{\pi}}{4}\;\frac{M(k_B
T_c)^{5/2}}{\sqrt{\gamma}\, \omega_z}\ ,
 \end{equation}
with $\zeta(5/2)\simeq 1.342$. The experimental results shown here
were obtained with $N=3\times 10^5$ rubidium atoms. This
corresponds to a critical temperature of $T_c=60$~nK for
$\Omega=\omega_\perp$, to be compared with $T_c=120$~nK for a
non-rotating gas.

Our experiments were performed in presence of radio-frequency
evaporation, which removes all atoms at a distance $r_\perp$
larger than $x_{\rm ev}=19\;\mu$m from the center (this
corresponds to an angular momentum value $m=x_{\rm
ev}^2/a_\perp^2\sim 200$). For $\Omega=\omega_\perp$ the well
depth is thus $U_0=\gamma x_{\rm ev}^4\simeq 60\,$nK, similar to
$k_B T_c$. Since the effective temperature $T$ in evaporative
cooling is a small fraction of the well depth (typically
$U_0/k_BT\sim 5$--$10$), the rotating gas is clearly in the
degenerate regime when $\Omega=\omega_\perp$.

\subsection{Observed vortex patterns}

\begin{figure}
\centerline{\includegraphics{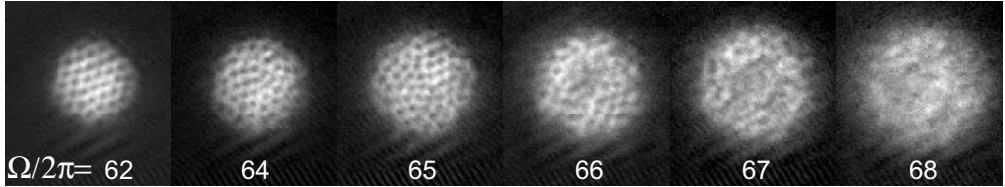}} \caption{\emph{Vortex
pattern in the fast rotation regime.} Pictures of the rubidium
condensate rotating in the quadratic+quartic potential, for
various stirring frequencies $\Omega/2\pi$. For these data
$\omega_\perp/2\pi=65$~Hz (pictures from \cite{Bretin04}). }
\label{rotcrit}
\end{figure}

We show in Fig. \ref{rotcrit} the images of the rotating gas as
the stirring frequency $\Omega$ is increased. For $\Omega
<\omega_\perp$, the vortex lattice is clearly visible. However
when $\Omega >\omega_\perp$ the visibility of the vortices
decreases and nearly vanishes for $\Omega=1.05\;\omega_\perp$
($=2\pi\times 68$~Hz).

The most plausible explanation of this effect is that the vortex
lines are still present, but strongly bent when $\Omega>
\omega_\perp$. This bending may occur because of the finite
temperature of the gas. A recent theoretical study
\cite{Aftalion03} seems to favor this hypothesis: when looking for
the ground state of the system using imaginary time evolution of
the Gross-Pitaevskii equation, it was found that much longer
imaginary times were required to reach a well ordered vortex
lattice for $\Omega >\omega_\perp$ than for
 $\Omega <\omega_\perp$.

\subsection{Transverse monopole mode}

The study of the normal modes of a Bose-Einstein condensate
generally provides insightful information about the system. In
order to gain some understanding of the fast rotation regime, we
have studied the transverse monopole (or breathing) mode for
various rotation frequencies of the condensate.

For a 2D gas at rest in an isotropic harmonic potential of
frequency $\omega$, this mode has a frequency $\omega_{\rm
mp}=2\omega$, which does not depend on the strength of the
interactions \cite{Pitaevskii97,Kagan97}. The state of the
condensate at time $t$ can be derived from the state at time 0 by
a simple scaling transform. The same result holds for a 3D gas
confined in an axisymmetric, cigar shaped potential
\cite{Chevy02}. In particular the frequency of the transverse
breathing mode is still $\omega_{\rm mp}=2\omega_\perp$
\cite{Stringari96}. For a rotating condensate one could have
naively expected that the frequency of the mode is changed to
$2\sqrt{\omega_\perp^2-\Omega^2}$ as a consequence of the
centrifugal potential. As shown in \cite{Cozzini03} this result is
not correct and the predicted frequency  is still $\omega_{\rm
mp}=2\omega_\perp$ for all rotation frequencies $\Omega$. This is
a striking example of the influence of the Coriolis force on the
system: even though it affects neither the BEC transition
temperature nor the cloud's equilibrium shape at $T=0$, it has a
strong impact on the condensate's normal modes. Note that the
result $\omega_{\rm mp}=2\omega_\perp$ holds for any 2-dimensional
gas with contact interactions confined in a harmonic potential
\cite{Pitaevskii97,Kagan97}, so that the monopole mode cannot be
used to monitor any phase transition -- like that to a strongly
correlated state.

We have studied experimentally this mode in the quadratic+quartic
potential described above \cite{Stock04}. We have checked that the
frequency $\omega_{\rm mp}$ remains at $2\omega_\perp$ for
rotation frequencies notably smaller than $\omega_\perp$, for
which the quartic term plays no significant role. When the
rotation frequency approaches $\omega_\perp$ we measure however a
small deviation from this value. This deviation increases with the
rotation frequency and reaches $\sim 10\%$ when $\Omega\sim
\omega_\perp$. This result is a consequence of the action of the
quartic potential and it can be accounted for by a simple analytic
model \cite{Stock04}. Due to this deviation the monopole frequency
might represent a sensitive tool to monitor the emergence of new
quantum phases of the rotating gas.

\begin{figure}
 \centerline{\includegraphics[height=8cm]{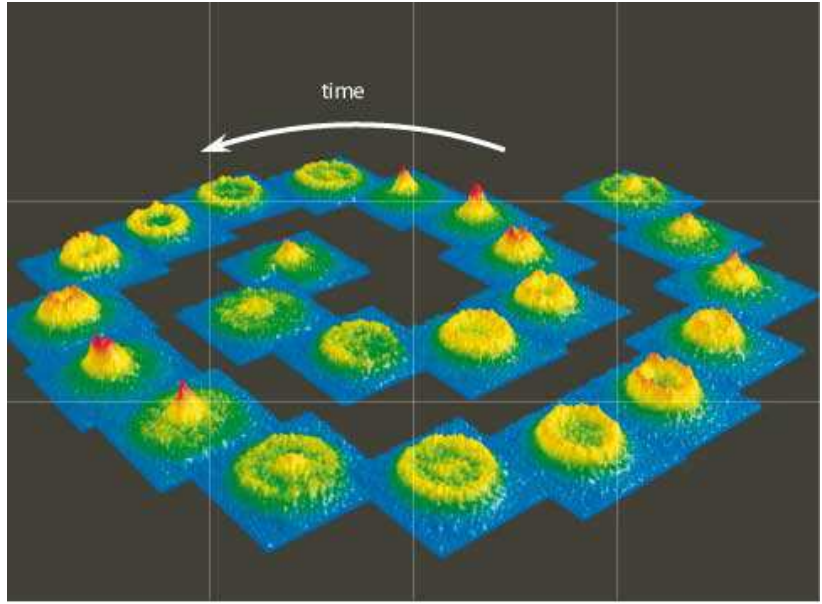}}
\caption{Transverse breathing mode of a condensate rotating at
$\Omega=1.05\; \omega_\perp$ ($\Omega/2\pi=68$~Hz). The condensate
is confined in the quadratic$+$quartic potential described in the
text. The structure of the mode corresponds to entering waves,
instead of a simple scaling transform as in the pure harmonic
case. The time interval between two successive pictures is 1 ms.
}\label{fig:breathing}
\end{figure}

A remarkable feature of the transverse monopole mode in the region
$\Omega\sim \omega_\perp$ is the time evolution of the density
profile of the cloud. Instead of being simply a scaling transform
as in the pure harmonic case, we observe a phenomenon of entering
waves (Fig.~\ref{fig:breathing}). This structure can be explained
by noticing that for $\Omega\sim \omega_\perp$ several $m=0$ modes
have a frequency close to $\omega_{\rm mp}$ \cite{Stock04}, so
that beating between them can lead to the observed phenomenon.

\section{Conclusions and perspectives}
\label{conclusion}

To conclude, the physics of a rotating Bose gas presents strong
analogies with several aspects of condensed matter physics:
superconductivity in large magnetic fields, Quantum Hall
phenomena, superfluidity and rotating bucket experiments. It has
already led to spectacular findings such as the possibility to
directly visualize the vortices and to observe their vibration
modes such as the Kelvin mode (oscillation of a vortex line) and
the Tkachenko mode (oscillation of the vortex lattice). However,
important aspects of the problem remain experimentally unexplored.
Let us briefly outline three lines of research that seem very
promising:
\begin{itemize}
\item
The possibility to generate quadratic $+$ quartic potential opens
the way to the nucleation of stable giant vortices. In a quadratic
potential a vortex with a topological charge larger than 1 is
unstable, and it can only be observed in a transient way
\cite{Engels03,Shin04}. When a quartic potential is present, this
instability may disappear and a giant vortex can be stabilized at
the center of the trap, possibly surrounded with singly-charged
vortices
\cite{Fetter01b,Kasamatsu02,Lundh02,Kavoulakis03,Fetter03,Aftalion03,Jackson04,Fischer03}

\item
The combination of rotation and optical lattices opens a very
interesting class of problems. If a 1D optical lattice is applied
along the axis of rotation, one obtains a stack of rotating
parallel layers. The structure of vortices in this system remains
to be investigated. One can qualitatively expect that the vortex
cores in neighboring layers will remain aligned if the tunneling
between layers is large enough, whereas they may decorrelate
otherwise \cite{Martikainen03}.

\item
When the rotation speed increases, the number of vortices $N_v$
also increases. When it becomes of the order of the particle
number $N$, one expects that the ground state of the gas will no
longer be well described by a mean-field approximation. Instead it
becomes a strongly correlated state, with a structure very similar
to those appearing in fractional quantum Hall effect
\cite{Cooper01,Paredes01,Sinova02,Reijnders02,Regnault03}. In
practice these type of states are expected to be observable only
for small particle numbers.

\end{itemize}

With such general lines of research still fully open, the next few
years should bring us a lot of novel and fascinating physics.

\vskip 1cm We thank  Amandine Aftalion, Yvan Castin, Christophe
Salomon, and Sandro Stringari for numerous useful discussions.
Z.H. acknowledges support from a Chateaubriand grant, and S.S.
from the Studienstiftung des deutschen Volkes, DAAD, and the
Research Training Network \emph{Cold Quantum Gases}
HPRN-CT-2000-00125. This work is supported by CNRS, Coll\`{e}ge de
France, R\'egion Ile de France, and DRED.



\begin{thebibliography}{99}

\bibitem{Cornell02} E. A. Cornell and  C. E. Wieman,
Rev. Mod. Phys. \textbf{74}, 875-893 (2002).

\bibitem{Ketterle02}
W. Ketterle, Rev. Mod. Phys. \textbf{74}, 1131-1151 (2002).

\bibitem{Pethickbook}
C.\ Pethick and H.\ Smith, \textit{Bose-Einstein condensation in
dilute Bose gases}, Cambridge University Press (2002).

\bibitem{Stringaribook} L. Pitaevskii and S. Stringari,
\textit{Bose-Einstein condensation}, Clarendon Press (2003).

\bibitem{Lifshitz}
E.M. Lifshitz and L. P. Pitaevskii, {\it Statistical Physics, Part
2}, chap. III (Butterworth-Heinemann, 1980).

\bibitem{Donnelly91}
R.J. Donnelly, {\it Quantized Vortices in Helium II}, (Cambridge,
1991), Chaps. 4 and 5.

\bibitem{Fetter01}
A. L. Fetter and A. A. Svidzinsky, J. Phys. Condens. Matter
\textbf{13}, R135 (2001).

\bibitem{Madison00}
K. W. Madison, F. Chevy, W. Wohlleben, and J. Dalibard, Phys. Rev.
Lett. {\bf 84}, 806, (2000).

\bibitem{Aboshaeer01}
J.R. Abo-Shaeer, C. Raman, J.M Vogels, and W. Ketterle, Science
\textbf{292}, 476 (2001);

\bibitem{Engels02}
P. Engels, I. Coddington, P. C. Haljan, and  E. A. Cornell, Phys.
Rev. Lett. \textbf{89}, 100403 (2002).

\bibitem{Rosenbusch02}
P. Rosenbusch, D. S. Petrov, S. Sinha, F. Chevy, V. Bretin, Y.
Castin, G. Shlyapnikov, and J. Dalibard, Phys. Rev. Lett.
\textbf{88}, 250403 (2002).

\bibitem{Feynman}
R. P. Feynman, in \emph{Progress in Low Temperature Physics}, vol.
1,  Chapter 2, C.J. Gorter Ed. (North-Holland, Amsterdam, 1955).

\bibitem{Cooper01}
N. R. Cooper, N. K. Wilkin, and J. M. F. Gunn, Phys. Rev. Lett.
\textbf{87}, 120405 (2001).

\bibitem{Paredes01}
B. Paredes, P. Fedichev, J. I. Cirac, and P. Zoller, Phys. Rev.
Lett. \textbf{87}, 010402 (2001).

\bibitem{Sinova02}
J. Sinova, C. B. Hanna, and A. H. MacDonald, Phys. Rev. Lett.
\textbf{89}, 030403 (2002).

\bibitem{Reijnders02}
J. W. Reijnders, F. J. M. van Lankvelt, K. Schoutens, and N. Read,
Phys. Rev. Lett. \textbf{89}, 120401 (2002)

\bibitem{Regnault03}
N. Regnault and T. Jolicoeur, Phys. Rev. Lett. \textbf{91}, 030402
(2003).

\bibitem{Schweikhard04b}
V. Schweikhard, I. Coddington, P. Engels, S. Tung, and  E. A.
Cornell, Phys. Rev. Lett. \textbf{93}, 210403 (2004).

\bibitem{Matthews99}
M. R. Matthews, B. P. Anderson, P. C. Haljan, D. S. Hall, C. E.
Wieman, and E. A. Cornell, Phys. Rev. Lett., \textbf{83}, 2498
(1999).

\bibitem{Leanhardt02}
A. E. Leanhardt, A. Görlitz, A. P. Chikkatur, D. Kielpinski, Y.
Shin, D. E. Pritchard, and W. Ketterle, Phys. Rev. Lett.
\textbf{89}, 190403 (2002).

\bibitem{Haljan01}
P.C. Haljan, I. Coddington, P. Engels, and E.A. Cornell, Phys.
Rev. Lett. \textbf{87}, 210403 (2001).

\bibitem{Hodby01}
E. Hodby, G. Hechenblaikner, S. A. Hopkins, O.M. Marag\'o, and C.
J. Foot, Phys. Rev. Lett. \textbf{88}, 010405 (2001).

\bibitem{Engels03}
P. Engels, I. Coddington, P.C. Haljan, V. Schweikhard, and E.A.
Cornell, Phys. Rev. Lett. \textbf{90}, 170405 (2003).

\bibitem{Caradoc99}
B. M. Caradoc-Davies, R. J. Ballagh, and K. Burnett, Phys. Rev.
Lett. \textbf{83}, 895-898 (1999).

\bibitem{Sinha01}
S. Sinha and Y. Castin, Phys. Rev. Lett. \textbf{87}, 190402
(2001).

\bibitem{Recati01}
A. Recati, F. Zambelli, and S. Stringari, Phys. Rev. Lett.
\textbf{86}, 377 (2001).

\bibitem{Madison01}
K. W. Madison, F. Chevy, V. Bretin, and J. Dalibard, Phys. Rev.
Lett. \textbf{86}, 4443 (2001).

\bibitem{Dalfovo01}
F. Dalfovo and S. Stringari, Phys. Rev. A \textbf{63}, 011601
(2001).

\bibitem{Linn01}
M. Linn, M. Niemeyer, and A. L. Fetter, Phys. Rev. A \textbf{64},
023602 (2001).

\bibitem{Aboshaeer02}
J. R. Abo-Shaeer, C. Raman, and W. Ketterle, Phys. Rev. Lett.
\textbf{88}, 070409 (2002).

\bibitem{Williams02}
J. E. Williams, E. Zaremba, B. Jackson, T. Nikuni, and A. Griffin,
Phys. Rev. Lett. \textbf{88}, 070401 (2002).

\bibitem{Simula02}
T. P. Simula, S. M. M. Virtanen, and M. M. Salomaa, Phys. Rev. A
\textbf{66}, 035601 (2002).

\bibitem{Tsubota02}
M. Tsubota, K. Kasamatsu, and M. Ueda, Phys. Rev. A \textbf{65},
023603 (2002).

\bibitem{Penckwitt02}
A. A. Penckwitt, R. J. Ballagh, and C. W. Gardiner Phys. Rev.
Lett. \textbf{89}, 260402 (2002).

\bibitem{Kasamatsu03}
K. Kasamatsu, M. Tsubota, and M. Ueda, Phys. Rev. A \textbf{67},
033610 (2003).

\bibitem{Lundh03}
E. Lundh, J.-P. Martikainen, and K.-A. Suominen Phys. Rev. A
\textbf{67}, 063604 (2003).

\bibitem{Lobo04}
C. Lobo, A. Sinatra, and Y. Castin, Phys. Rev. Lett. \textbf{92},
020403 (2004).

\bibitem{Duine04}
R. A. Duine, B. W. A. Leurs, and H. T. C. Stoof,  Phys. Rev. A
\textbf{69}, 053623 (2004).

\bibitem{Chevy00}
F. Chevy, K. W. Madison, and J. Dalibard, Phys. Rev. Lett. {\bf
85}, 2223 (2000).

\bibitem{Zambelli98}
F. Zambelli and S. Stringari, Phys. Rev. Lett. {\bf 81}, 1754
(1998).

\bibitem{Inouye01}
S. Inouye, S. Gupta, T. Rosenband, A. P. Chikkatur, A. Görlitz, T.
L. Gustavson, A. E. Leanhardt, D. E. Pritchard, and W. Ketterle,
Phys. Rev. Lett. \textbf{87}, 080402 (2001).

\bibitem{Chevy01}
F. Chevy, K. W. Madison, V. Bretin, and J. Dalibard, Phys. Rev. A
\textbf{64}, 031601(R) (2001).

\bibitem{Rosenbusch02b}
P. Rosenbusch, V. Bretin, and J. Dalibard, Phys. Rev. Lett.
\textbf{89}, 200403 (2002).

\bibitem{Svidzinsky00}
A. A. Svidzinsky and A. L. Fetter, Phys. Rev. A {\bf 62}, 063617
(2000).

\bibitem{Feder01}
D. L. Feder, A. A. Svidzinsky, A. L. Fetter, and C. W. Clark,
Phys. Rev. Lett. \textbf{86}, 564 (2001).

\bibitem{Garcia01a}
J.J. Garc\'{\i}a-Ripoll and V.M. P\'erez-Garc\'{\i}a, Phys. Rev. A
\textbf{63}, 041603 (2001).

\bibitem{Garcia01b}
J.J. Garc\'{\i}a-Ripoll and V.M. P\'erez-Garc\'{\i}a, Phys. Rev. A
\textbf{64}, 053611 (2001).

\bibitem{Aftalion01}
A. Aftalion and T. Riviere, Phys. Rev. A \textbf{64}, 043611
(2001).

\bibitem{Modugno03}
M. Modugno, L. Pricoupenko, and Y. Castin, Eur. Phys. J. D
\textbf{22}, 235 (2003).

\bibitem{Fedichev99}
P. O. Fedichev and G. V. Shlyapnikov, Phys. Rev. A \textbf{60},
R1779 (1999).

\bibitem{Anderson00}
B. P. Anderson, P. C. Haljan, C. E. Wieman, and E. A. Cornell,
Phys. Rev. Lett. \textbf{85}, 2857-2860 (2000).

\bibitem{Stringari01}
S. Stringari, Phys. Rev. Lett. \textbf{86}, 4725-4728 (2001).

\bibitem{Hodby03}
E. Hodby, S. A. Hopkins, G. Hechenblaikner, N. L. Smith, and C. J.
Foot, Phys. Rev. Lett. \textbf{91}, 090403 (2003).

\bibitem{Bretin03}
V. Bretin, P. Rosenbusch, F. Chevy, G. V. Shlyapnikov, and J.
Dalibard, Phys. Rev. Lett. \textbf{90}, 100403 (2003).

\bibitem{Mizushima03}
T. Mizushima, M. Ichioda, and K. Machida, Phys. Rev. Lett.
\textbf{90}, 180401 (2003).

\bibitem{Chevy03}
F. Chevy and S. Stringari, Phys. Rev. A \textbf{68}, 053601
(2003).

\bibitem{Duine03}
R. A. Duine and H. T. C. Stoof, Phys. Rev. Lett. \textbf{91},
150405 (2003).

\bibitem{Fetter04}
A. L. Fetter, Phys. Rev. A \textbf{69}, 043617 (2004).

\bibitem{Girvin84}
S. M. Girvin and T. Jach, Phys. Rev. B \textbf{29}, 5617 (1984).

\bibitem{Rokhsar99}
D. A. Butts and D. S. Rokhsar, Nature \textbf{397}, 327 (1999).

\bibitem{Ho01}
T. L. Ho, Phys. Rev. Lett. {\bf 87} 060403 (2001).

\bibitem{Stringari99}
S. Stringari, Phys. Rev. Lett. \textbf{82}, 4371-4375 (1999)

\bibitem{Cozzini03}
M. Cozzini and S. Stringari, Phys. Rev. A \textbf{67}, 041602
(2003).

\bibitem{Aftalion04}
A. Aftalion, X. Blanc, and J. Dalibard, cond-mat/0410665; to
appear in Phys. Rev. A.

\bibitem{Baym04a}
G. Baym and C. J. Pethick, Phys. Rev. A, {\bf 69} (2004).

\bibitem{Watanabe04}
G. Watanabe, G. Baym and C. J. Pethick, Phys. Rev. Lett.
\textbf{93}, 190401 (2004)

\bibitem{Cooper04}
N. R. Cooper, S. Komineas and N. Read, Phys. Rev. A \textbf{70},
033604 (2004).

\bibitem{Kleiner64}
W. H. Kleiner, L. M. Roth and S. H. Autler, Phys. Rev. {\bf 133},
A1226, (1964).

\bibitem{Schweikhard04}
V. Schweikhard, I. Coddington, P. Engels, V. P. Mogendorff, and E.
A. Cornell, Phys. Rev. Lett. \textbf{92}, 040404 (2004).

\bibitem{Coddington04}
I. Coddington, P. C. Haljan, P. Engels, V. Schweikhard, S. Tung,
E. A. Cornell, cond-mat/0405240.

\bibitem{Raman01}
C. Raman, J.R. Abo-Shaeer, J.M. Vogels, K.Xu, and W. Ketterle,
Phys. Rev. Lett. {\bf 87}, 210402 (2001).

\bibitem{Castin99}
Y. Castin and R. Dum, Eur. Phys. J. D, \textbf{7}, 399-412 (1999).

\bibitem{Sheehy04}
D. E. Sheehy and L. Radzihovsky, cond-mat/0402637 and
cond-mat/0406205.

\bibitem{Fischer03}
U.R. Fischer and G. Baym, Phys. Rev. Lett. \textbf{90}, 140402
(2003).

\bibitem{Anglin02}
J. R. Anglin and M. Crescimanno, cond-mat/0210063.

\bibitem{Coddington03}
I. Coddington, P. Engels, V. Schweikhard, and E. A. Cornell, Phys.
Rev. Lett. \textbf{91}, 100402 (2003).

\bibitem{Baym03}
G. Baym, Phys. Rev. Lett. \textbf{91}, 110402 (2003).

\bibitem{Choi03}
S. Choi, L. O. Baksmaty, S. J. Woo, and N. P. Bigelow, Phys. Rev.
A. \textbf{68}, 031605 (2003).

\bibitem{Simula04}
T. P. Simula, A. A. Penckwitt, and R. J. Ballagh Phys. Rev. Lett.
92, 060401 (2004)

\bibitem{Mizushima04}
T. Mizushima, Y. Kawaguchi, K. Machida, T. Ohmi, T. Isoshima, and
M. M. Salomaa, Phys. Rev. Lett. \textbf{92}, 060407 (2004).

\bibitem{Gifford04}
S. A. Gifford and G. Baym, cond-mat/0405182.

\bibitem{Cozzini04}
M. Cozzini, L. P. Pitaevskii, and S. Stringari, Phys. Rev. Lett.
\textbf{92}, 220401 (2004).

\bibitem{Fetter01b}
A.L. Fetter, Phys. Rev. A \textbf{64}, 063608 (2001).

\bibitem{Kasamatsu02}
K. Kasamatsu, M. Tsubota, and M. Ueda, Phys. Rev. A \textbf{66},
053606 (2002).

\bibitem{Lundh02} E. Lundh, Phys. Rev. A \textbf{65}, 043604 (2002).

\bibitem{Kavoulakis03}
G.M. Kavoulakis and G. Baym, New Jour. Phys. \textbf{5}, 51.1
(2003).

\bibitem{Fetter03}
A. L. Fetter, Phys. Rev. A \textbf{68}, 063617 (2003).

\bibitem{Aftalion03}
A. Aftalion and I. Danaila, Phys. Rev. A \textbf{69}, 033608
(2004).

\bibitem{Jackson04}
 A. D. Jackson, G. M. Kavoulakis, and E. Lundh, Phys. Rev. A \textbf{69},
053619 (2004); see also A. D. Jackson and G. M. Kavoulakis,
cond-mat/0311066.

\bibitem{Fetter04b}
A. L. Fetter, B. Jackson, and S. Stringari, cond-mat/0407119.

\bibitem{Bretin04}
V. Bretin, S. Stock, Y. Seurin, and J. Dalibard Phys. Rev. Lett.
\textbf{92}, 050403 (2004).

\bibitem{Pitaevskii97}
L. P. Pitaevskii and A. Rosch, Phys. Rev. A \textbf{55} R853
(1997).

\bibitem{Kagan97}
Yu. Kagan, E.L. Surkov, and G.V. Shlyapnikov, Phys. Rev. A
\textbf{54}, R1753 (1996).

\bibitem{Chevy02}
F. Chevy, V. Bretin, P. Rosenbusch, K. W. Madison, and J.
Dalibard, Phys. Rev. Lett. \textbf{88}, 250402 (2002).

\bibitem{Stringari96}
S. Stringari, Phys. Rev. Lett. {\bf 77}, 2360 (1996).

\bibitem{Stock04}
S. Stock, V. Bretin, F. Chevy and J. Dalibard, Europhys. Lett.
\textbf{65}, 594 (2004).

\bibitem{Shin04}
Y. Shin, M. Saba, M. Vengalattore, T. A. Pasquini, C. Sanner, A.
E. Leanhardt, M. Prentiss, D. E. Pritchard, and W. Ketterle, Phys.
Rev. Lett. \textbf{93}, 160406 (2004).

\bibitem{Martikainen03}
J.-P. Martikainen and H. T. C. Stoof, Phys. Rev. Lett.
\textbf{91}, 240403 (2003); Phys. Rev. A \textbf{69}, 053617
(2004); Phys. Rev. Lett. \textbf{93}, 070402 (2004).

\end{thebibliography}
\end{document}